  \documentclass[aps,article,twocolumn,superscriptaddress]{revtex4-1}
\usepackage{graphics} 
\usepackage{mathtools}
  \usepackage{lineno}

\begin{document}

\title{Sequential steady state co-rotating dust vortices in sheared streaming plasma}
\author{Modhuchandra Laishram}\email{modhu@ustc.edu.cn}
\address{CAS Key Laboratory of Geospace Environment and Department of Engineering 
and Applied Physics, University of Science and Technology of China, Hefei 230026,
China}

  \author{Devendra Sharma}
\address{Institute for Plasma Research, HBNI, Bhat, Gandhinagar 382428, India}

  \author{Ping Zhu}\email{pzhu@ustc.edu.cn}
\address{CAS Key Laboratory of Geospace Environment and Department of Engineering 
and Applied Physics, University of Science and Technology of China, Hefei 230026,
China}
\address{KTX Laboratory and Department of Engineering and Applied Physics, 
University of Science and Technology of China, Hefei 230026, China}
\address{Department of Engineering Physics, University of Wisconsin-Madison, 
Madison, Wisconsin 53706, USA}

\date{\today}

\begin{abstract}
The 2D hydrodynamic model for a dust cloud confined in an axisymmetric toroidal 
system volumetrically driven by an unbounded streaming plasma  
is further extended systematically for different aspect-ratio of the 
bounded dust domain and a wide range of the kinematic viscosity. 
This work has demonstrated the interplay between inertial and diffusive transport 
processes for the structural changes of steady dust flow 
from symmetric into asymmetric nature in higher Reynolds number (Re) 
regimes where flow streamlines turn more circular and 
the structural bifurcation takes place through a threshold parameter. 
In agreement with many experimental observations, the steady vortex 
structure in highly nonlinear (i.e., high Re) regime is characterized by 
the critical transition into a new self-similar multiple co-rotating 
vortices, along with circular core region of single 
characteristics size and surrounded by strongly 
sheared layers filled with weak vortices near the boundaries. 
It is further revealed that the core region persists for a wide 
range of system parameters in the nonlinear regime 
and its characteristic size is mainly determined by the smallest distance 
between the confining boundaries. The threshold parameter, 
the vortex size, the strength, and the number of 
the self-similar co-rotating vortices mainly depend on the 
aspect-ratio of the bounded dust domain.
These nonlinear solutions provide insight into the phenomena of 
the structural transition and coexistence of self-similar 
steady co-rotating vortices in dusty plasma experiments as well as many 
relevant complex driven-dissipative natural flow systems.
\end{abstract}
\maketitle
\section{Introduction}\label{introduction}
Vortices are fundamental constituents of turbulence flow and collective 
dynamics in various driven-dissipative systems~\cite{PhysRevE.95.033204,mcwilliams_1990,
PhysRevLett.112.115002,MEUNIER2005431,Nivedita}. They have been 
observed in different shapes, sizes, strengths, and directions in varieties 
of laboratory dusty plasma experiments and biophysical complex flow 
systems~\cite{doi:10.1063/1.4929916,1367-2630-5-1-366,doi:10.1063/1.4941973,
doi:10.1063/1.4977454,doi:10.1063/1.5019364,doi:10.1002/ctpp.201700039,Nivedita}. 
And the existence of adjacent multiple counter-rotating vortices is intuitively
explained by the shear nature of driving mechanisms~\cite{PhysRevE.91.063110,
doi:10.1063/1.4929916}. However, the appearance of self-similar co-rotating 
dust vortices reported in recent experiments with a background 
plasma flow of monotonic shear is rather counter-intuitive since a sharp dust 
flow shear layer is essentially present between two spatially adjacent 
co-rotating vortices, usually unexpected in steady fluid flow 
equilibria~\cite{doi:10.1063/1.4977454,
doi:10.1063/1.4941973,doi:10.1002/ctpp.201700039}. In such cases, 
the physics of single-fluid multiple vortices (like the 
Taylor-Couette vortices~\cite{Taylor289,Stuart}) may not be directly 
applicable to the volumetrically driven multiple-fluid cases such as those 
found in dusty plasma and biophysical complex 
systems~\cite{doi:10.1063/1.5019364,doi:10.1063/1.4941973,Nivedita}. 

The 2D hydrodynamic approach to the analysis of the dust vortex flow in 
plasma has been presented in a series of previous work~\cite{PhysRevE.91.063110,
PhysRevE.95.033204,doi:10.1063/1.4887003,doi:10.1063/1.5045772}, 
in both linear and nonlinear regimes with the motivation to 
interpret the observations in laboratory dusty 
plasmas and those in micro-gravity international 
space station (ISS)~\cite{doi:10.1063/1.4929916,1367-2630-5-1-366}. 
The nonlinear analysis~\cite{PhysRevE.95.033204} briefly addresses 
the possible accessibility to secondary co-rotating vortices in the 
solutions it recovered in a square domain (AR=1) of dust confinement where the 
ratio of secondary to primary vortex remains limited to be $1: (\sqrt{2}-1)$. 
The possibility of self-similar co-rotating vortices was also 
discussed in the previous work predicting in advance the applicability of 
such solutions to some experiments where the recovery of co-rotating vortices 
are being explored~\cite{doi:10.1063/1.4977454,doi:10.1063/1.5019364}. 
 
A partial freedom, allowed by the multiple equal-sized co-rotating vortex 
formations, from the bounded geometry of arbitrary shape is the subject of 
this paper. We discuss how an alternative convective mechanism facilitates 
the momentum transports and produces a rather
non-intuitive and highly sheared arrangement of multiple adjacent 
co-rotating circular vortex flows observed in many recent experiments.
For example, in the dusty plasma experimental 
work, Kil-Byoung Chai {\it et~al.}~(2016)~\cite{doi:10.1063/1.4941973} reported 
recovery of two adjacent co-rotating and counter-rotating poloidal vortices due 
to the presence of ion density gradient and the gradient of ion ambipolar 
velocity. They also observed the structural transition of the dust cloud into 
multiple vortices when the plasma density exceeds a critical value ${n_i}^*$. 
Further, in the series of recent experiments carried out by 
M. Choudhary {\it et~al.}~(2018)~\cite{doi:10.1063/1.4977454,doi:10.1063/1.5019364}, 
who has analyzed dust dynamics in a domain of higher aspect-ratio 
$L_z/L_r \approx$3~to~5, indeed observed the structural transition 
from a single to co-rotating dust vortices of equal 
size by changing the applied discharge 
power through a threshold value ${P}^*$. Further, in some cases, they also 
observed variation in the number of co-rotating vortices at different cross-sections 
of the elongated 3D dust cloud in an inhomogeneous rf-plasma. 
In addition to this, apart from dusty plasma i.e., in the biophysical complex 
fluid flows, Nivedita {et~al.~(2017)}~\cite{Nivedita} have 
observed similar kind of structural changes and transition into multiple co-rotating 
vortices through a threshold in Dean number $D^*$ in microfluidics systems such as 
spiral microchannels for cell sorting and micromixing devices. 
Here, D is a modified form of Re taking into account of aspect-ratio 
of the flow system. This work demonstrates a critical phenomena in term of an 
instability arising from the imbalance between flow pressure and velocity
gradient near the boundary. 

Although these observations have been reported in part, 
the actual physical interpretations for the steady vortex structural changes 
why the flow trajectory turns more circular with increasing the convective transport 
is not well understood.
Then, the physics of the persistence of self-similar multiple 
co-rotating vortices, and the roles of aspect-ratio of the flow domain 
in the laboratory dusty plasma experiments
are not discussed explicitly. 
Further, the phenomena of the steady dust vortex changes in 
characteristic size, numbers, strength, and direction of the self-similar multiple 
vortices remain unclear in the experiments, but the observations indicate that 
a combination of the dynamic regime and the aspect-ratio of the flow domain 
plays an important role in determining the vortex size and overall flow 
structure. Therefore, in this present work, we emphasize mostly on the effect of 
aspect-ratio, by extending the previous nonlinear analysis in the fixed domain
of aspect-ratio unity (AR=1)~\cite{PhysRevE.95.033204} to different 
aspect-ratio of the flow domain and varying a wide range of kinematic viscosities. 
The present work provides a more complete analysis and characterization of 
dust vortex dynamics where a sequence of self-similar co-rotating vortices are 
obtained in the domain of different aspect-ratios (AR=$2,3$) relevant to recent 
experiments. 

The manuscript is organized as follows. In Sec.~\ref{formulation}, 
we recall the 2D hydrodynamic model for the 
bounded dust flow in streaming plasma from the previous 
analysis~\cite{doi:10.1063/1.5045772}, which is followed by 
the characterization of system parameters for the dynamic regime in 
Sec.~\ref{analysis}. Then, the qualitative 
effects of domain aspect-ratio and kinematic viscosity (or Re) 
are discussed in Sec.~\ref{AR-effects}. 
Further, we examine some of the recent results in dusty plasma experiments 
and other complex fluids system using our model in Sec.~\ref{experimental_obsrv}. 
Summary and conclusion is presented in Sec.~\ref{conclusion}.

\begin{figure}\centering
\includegraphics{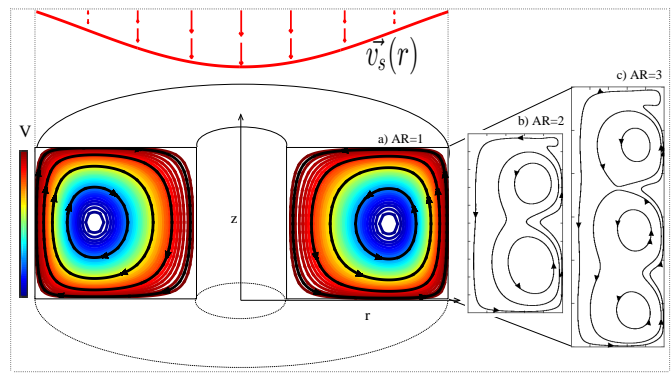}
\caption{Schematic representation of the dust cloud confined in an azimuthally 
symmetric torus by the effective potential $V(r,z)$ and driven by the 
shear ions field $v_s(r,x)$ interspersing through the bounded dust cloud, 
also demonstrating the steady-state dust vortex structure 
in nonlinear regimes for various aspect-ratio (AR = $L_{z}/L_{r}$) of the 
confined domain.  
 \label{schematic}}
\end{figure}
\section{2D nonlinear hydrodynamic model}
\label{formulation}
Similar to the previous analysis~\cite{PhysRevE.95.033204,doi:10.1063/1.5045772}, 
we consider a nearly incompressible dust fluid confined in an axisymmetric 
toroidal system along with an unbounded flowing plasma. 
The geometry of bounded dust fluid is very similar to that of 
recent experiments where a toroidal dust cloud flowing in the poloidal 
direction in a glow discharge plasma~\cite{doi:10.1063/1.4929916}. 
For computational simplicity, the poloidal cross-section 
of the axisymmetric torus is simplified as a rectangular domain as illustrated 
schematically in Fig~\ref{schematic}. The dust fluid is confined by an 
effective potential $V(r,z)$ in the toroidal region 
where $0<r/L_{r}<1$, $-1<z/L_{z}<1$, and symmetric along $\hat{\phi}$, 
which can vary the size (AR = $L_{z}/L_{r}$) of dust confining domain, where the  
boundary serves as the equipotential line for the perfect confinement.

Then, using the stream function-vorticity formulation discussed in 
the previous work~\cite{PhysRevE.95.033204,doi:10.1063/1.5045772}, the dynamics of the 
dust flow that is 
incompressible, 
isothermal, 
and has a finite viscosity, is governed by the modified 
Navier-Stokes equations where the sheared ion 
drag and the friction from the stationary background neutral fluid
can suitably be accounted as non-conservative momentum source and sink as 
follows~\cite{landau2013fluid},
%
%
%
\begin{eqnarray} 
\nabla^2\psi&=&-\omega,
\label{streamfunction-equation}\\
({\bf u} \cdot \nabla) {\omega}&=&\mu\nabla^2\omega -\xi(\omega-\omega_s) -\nu\omega.  
\label{vorticity-equation}
\end{eqnarray}
Here, $\psi\hat{\phi}$ is the stream function giving incompressible dust flow  
velocity ${\bf u}=\nabla\times\psi\hat{\phi}$ in the 2D $rz$-plane, 
$\omega\hat{\phi} =\nabla\times{\bf u}$ is 
the corresponding dust vorticity, $\mu$ is the dust kinematic viscosity, 
$\xi$ is the coefficient of ion drag acting on the dust, and $\nu$ is the coefficient 
of friction generated by the stationary neutral 
fluid~\cite{PhysRevLett.68.313,PhysRevE.66.046414,PhysRevLett.92.205007}.
Further, $\omega_{s}$ is the external vorticity source from the unbounded 
background flows. In a real system, any non-zero shear field such as 
$\nabla\times u_{i(n)}$,
($\nabla Q_d\times{\bf E}$), 
($\nabla u_{i(n)} \times \nabla n_{i(n)} $) and 
($\nabla T_{i(n)} \times \nabla n_{i(n)} $) of the complex background flow 
and their combinations can be driving mechanisms and contribute 
to the source of vorticity {$\omega_s$}, because the 
coefficient $\xi,~\nu$, and $\mu$ depends on the state of the system
\cite{PhysRevLett.68.313,PhysRevE.66.046414,
PhysRevLett.92.205007,doi:10.1063/1.5020416,doi:10.1063/1.4941973}. 
Here, $Q_d$ is dust charge, E is effective electric field along the streaming ions, and 
the subscript $i(n)$ represents the background ion (neutral).
The overall combination of charged dust and background plasma is 
quasi-neutral and the highly mobile background electrons and ions are in thermal 
equilibrium even though it interacts and intersperses through the confined 
dust fluids. Therefore, the steady equilibrium shear ions vorticity 
field $\omega_{s}$ in Eqn~(\ref{vorticity-equation}) 
is valid for representing the dominant ion drag interaction with the bounded 
dust dynamics~\cite{doi:10.1063/1.5045772}.

\section{Characterization of the bounded dust fluid system}
\label{analysis}
%
Now, the set of equations (\ref{streamfunction-equation})-(\ref{vorticity-equation}), 
represent the steady state of the driven dust fluid for any dynamical regime 
defined by the system parameters $\xi,~\nu$, $\mu$, 
the driving field $\omega_s$, and the boundary conditions. 
A perfect slip, partial slip, and no slip 
are some of the very common boundary conditions 
encountered in dusty plasma. 
These equations are solved subjected to similar boundary conditions and 
numerical technique adopted from the previous 
work~\cite{PhysRevE.95.033204,doi:10.1063/1.5045772}. 

\begin{figure}
\includegraphics{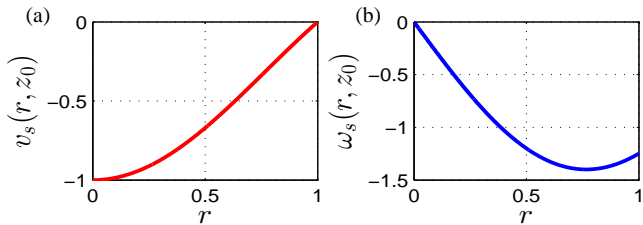}
\caption{Profiles of the external (a) driver velocity $v_s$, 
and the corresponding 
(b) driver vorticity $\omega_{s}$ which are uniform 
along axial $\hat{z}$ 
and azimuthal $\hat{\phi}$ directions.
\label{source}}
\end{figure}

For the main driving field, i.e., the vertically streaming sheared ions directed 
along $-\hat z$ ($v_r=0$) is considered, motivated by various 
experiments~\cite{doi:10.1063/1.4929916,doi:10.1063/1.5019364}.
The radial profile of the streaming ions have a radial shear nature of the 
single natural mode of the nonplanar system as given in 
Eqn.~(\ref{vz_bessel}), (also plotted in Fig~\ref{source}), 
such that it maintains maximum velocity at 
the inner radial boundary $(r=r_1\approx 0)$ and minimum velocity 
at the external confining boundary $(r=L_r)$ of the bounded domain. 
{\small
\begin{eqnarray}
v_{z}=U_{c}+U_{0}J_0\left(\alpha_m \frac{r-r_1}{L_{r}-r_1}\right).
\label{vz_bessel}
\end{eqnarray}
}
Here $U_{c}$ 
represents an offset, and $U_{0}$ is the strength of
radial variation of the ions flow. 
The radial mode number $\alpha_m$ represents the zeros of 
corresponding ion velocity profile coinciding with the external 
boundary location $L_r$. The entire analysis is done using 
the same driver velocity $v_{z}(r,z)$ and 
its corresponding vorticity $\omega_{s}(r,z)=\nabla\times v_{z}$ 
profile as in the previous work~\cite{PhysRevE.95.033204}.

Now, for the system parameters, the solutions presented here 
correspond to a typical laboratory glow discharge argon plasma 
with micron size dust, along with parameters 
$n \simeq 10^{9}$ cm$^{-3}$, $T_{e}\simeq 3 eV$, $T_{i}\simeq 1 eV$, 
and shear ions are streaming with the value $U_{0}$ 
equivalent to the fraction of the ion acoustic velocity 
$c_{s}=\sqrt{T_{e}/m_{i}}$.  Then, using the radial width of the 
confined domain $ L_r$ and steaming shear ions 
velocity strength $U_{0}$ as the ideal normalization units, 
the value of ion drag coefficient and neutral collision 
frequency can be estimated as $\xi\sim 10^{-4}~ U_{0}/L_r$ and 
$\nu\sim 10^{-3}~ U_{0}/L_r$, 
respectively~\cite{PhysRevLett.68.313,PhysRevE.66.046414,
PhysRevLett.92.205007}.
Further, for a typical system, $L_r \sim 10$ cm, the range of 
kinematic viscosity $\mu$ can similarly be chosen to be 
$\mu\sim 6\times 10^{-4} ~ U_{0}L_r$ 
which corresponds to small Reynolds numbers (Re $\simeq 1$) of the 
dust flow consistent with the linear viscous 
regime~\cite{PhysRevLett.88.065002,PhysRevE.91.063110,PhysRevE.95.033204}. 
\section{Flow structure dependence on aspect-ratio of the bounded domain
\label{AR-effects}}
In the previous analytical work~\cite{PhysRevE.91.063110}, it has 
demonstrated that the vortex structure in the low velocity or linear regimes 
is mainly determined by the scale introduced by the external driving fields 
and the geometry of the bounded domain. However, in the high-velocity nonlinear 
regime~\cite{doi:10.1063/1.5045772,PhysRevE.95.033204}, the steady flow structure 
is again influenced mainly by the dynamical regime in addition to the 
scales introduced by the geometry of the bounded domain. Thus, the flow structural 
change continuously in response to the increase in strength of advective transport 
process $(\bf{u} \cdot \nabla) \bf{\omega}$  relative to the diffusive transport 
process $\mu\nabla^2 {\bf \omega}$. Along with this structural changes in the nonlinear 
regime, the nonlinear structure bifurcation takes place through a threshold 
value $\mu*$~\cite{PhysRevE.95.033204}, and the flow turns into a more complex 
self-organized structure consisting of multiple vortices of 
varying shape, size, strength, and direction. 

In terms of scales available in the flow system, the diffusive transport process 
mainly depends on the characteristic length across the flows $L_{\perp}$, whereas 
convective transport depends on characteristic length $L_{\|} (\approx u/u')$ 
along flow streamlines. Therefore, the nonlinear structural changes with 
diffusive coefficient $\mu$ can also be described in terms of the variation in 
characteristic length $L_{\perp}$ relative to the $L_{\|}$ and vice versa, which 
are correlated to the changes in aspect-ratio ($AR=L_{z}/L_{r} $) of the bounded 
flow domain. Thus, the aspect-ratio of the bounded flow has an important role in 
determining the steady-state flow structure. The dependence of bounded flow 
characteristics on the domain aspect-ratio can also be visualized in a 
simple way as follow. For the volumetrically driven system of 
aspect-ratio AR($=L_z/L_r$) with a driver having the fixed monotonic shear 
scale $L_r$ and uniform dissipative neutrals at the background, 
the vertically streaming ions can intersperse through the bounded fluid for 
maximum range up to $L_z$. Therefore, for higher AR ($L_z \gg L_r$), 
the shear ions flow can interact or 
transfer relatively larger momentum to the confined dust in comparison to 
the case of small aspect-ratio ($L_z \le L_r$), and hence the dust cloud 
retains more and more momentum from the source ions. As a consequence, 
the structural changes with system parameters are very different in various 
aspect-ratio of the bounded flow domain.  
In the following, we examine the steady state dust flow 
structural changes for a wide range of kinematic viscosity $\mu$ in different 
aspect-ratio of the bounded dust flow domain. 
\subsection{Steady dust flow structure in the confined domain of 
aspect-ratio $AR =2$ \label{aspect_ratio_2}}
\begin{figure}
 \centering
      \includegraphics[width=0.5\textwidth]{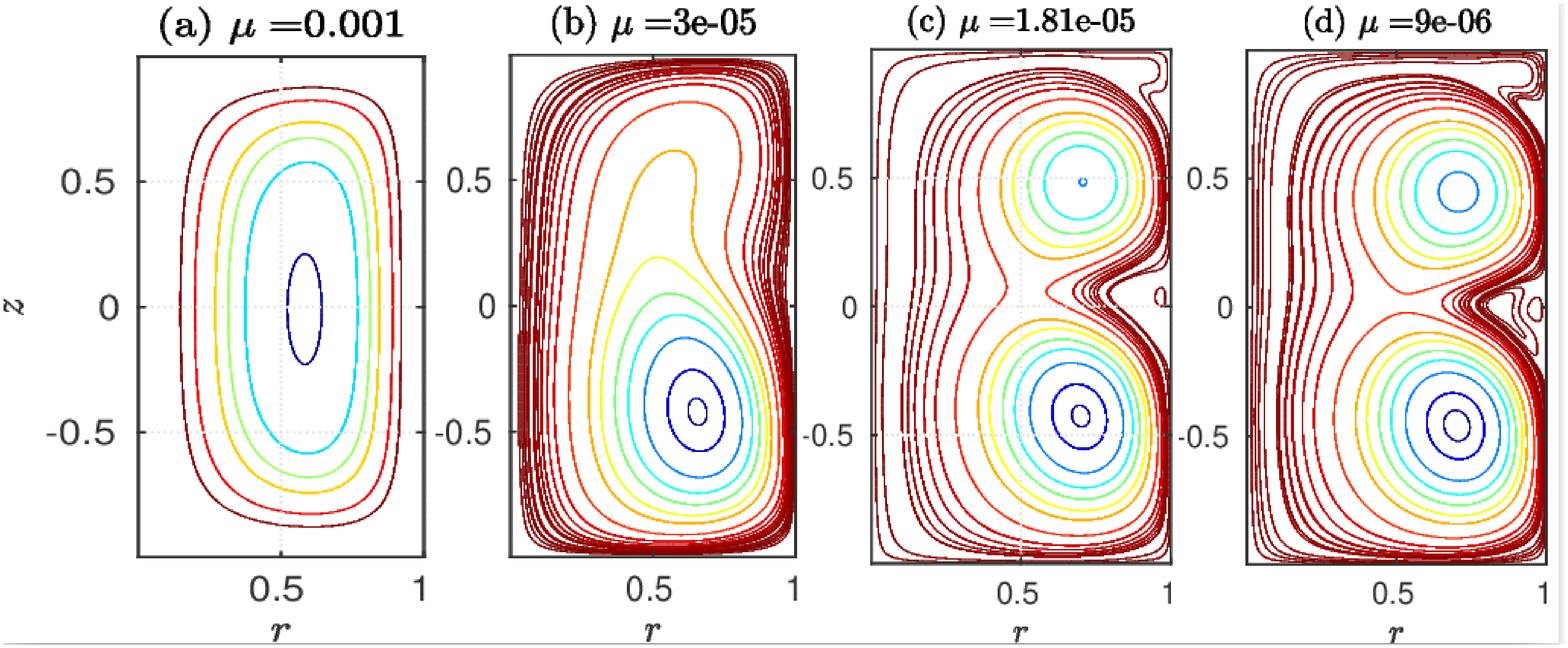}\\
     \includegraphics{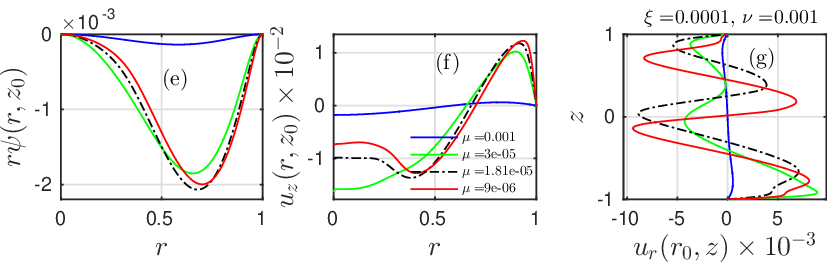}
 \caption{\small Streamlines for the steady bounded dust flow in $r$-$z$ 
 cross-section of $AR = 2$ for varying 
 $(a)~\mu = 10^{-3} U_{0}L_{r}$, 
 $(b)~\mu = 3\times 10^{-5}U_{0}L_{r}$
 $(c)~\mu = 1.8\times 10^{-5}U_{0}L_{r}$
 and $(d)~\mu = 9\times 10^{-6} U_{0}L_{r} $ respectively 
having fixed other system parameters $ \xi = 10^{-4} U_{0}/L_{r}$, 
$\nu =  10^{-3} U_{0}/L_{r}$. 
The corresponding cross-section profiles for $~(e)~ r\psi(r,z_0)$, 
$~(f)~ u_{z}(r,z_0)$ and $~(g)~ u_{r}(r_0,z)$ passing through the center of 
the primary vortex. 
 \label{ar-2}
}
\end{figure}

 A series of steady state dust flow structure in a confining domain of 
aspect-ratio $AR =2$ are shown in Fig.~\ref{ar-2}, for the wide range of 
kinematic viscosity $\mu$ evolving from linear to nonlinear regimes 
without changing any other system parameters.
%
As discussed in the recent work~\cite{doi:10.1063/1.5045772}, 
the flow structure in the confined 
domain in the case of highly viscous regime $\mu \sim 10^{-3} U_{0}L$ 
(or Re $\le 10$) is symmetric and elongated circulations 
as shown in Fig.~\ref{ar-2}$(a)$.  When the diffusive coefficient reduces 
to $\mu \sim 3\times10^{-5} U_{0}L$ (or 10 $\le$ Re $\le$ 40),
the vortex structure turns into asymmetric about the centerline, 
the center of circulation drifted axially downward and 
radially outward to a new position in the new state as 
shown in Fig.~\ref{ar-2}$(b)$. 
The observed structural changes with varying $\mu$ from symmetric to 
asymmetric developing highly shear layers near 
the boundary is the beginning of the nonlinear effect 
arises in the flow~\cite{mitic2008convective,Hall2016ASO}. 
Then the nonlinear effect becomes more strengthened with further 
reduces in the diffusive coefficient up to $\mu \sim 10^{-5} U_{0}L$ or 
less ( i.e., increase in Re $\ge 80$), and as a consequence, the nonlinear
structural bifurcation takes place through a critical value 
$\mu^* \approx 1.81\times10^{-5} U_{0}L_{r}$ giving multiple co-rotating 
vortices of the bounded dust cloud in the flow domain~\cite{PhysRevE.95.033204}.

Now, in a steady circulating flow with fixed sources and sinks, 
the fluid element retains more and more momentum along the closed paths with 
a decrease in the diffusive transport process. Therefore, the flow trajectory turned 
more circular since the circular path allows to keep maximum 
angular velocity $(\approx 2\omega$) without a significant change in 
the angular momentum of the whole system. Thus, the new vortex structure 
in the highly angular velocity regime is characterized by the emergence of a 
circular core region of single scale i.e., the characteristic size of the 
circular core ($s$), on which the Prandlt-Batchelor theorem is 
satisfied ($\frac{\partial\omega}{\partial\psi}\approx 0$ means 
$\mu\oint \nabla^2 u\cdot d{\bf l}\approx0$)
~\cite{PhysRevE.95.033204,batchelor_1956}. This means the steady core region 
of uniform vorticity surrounded by highly shear collar layer is free from viscous 
dissipation, and hence it persists for a wide range of system 
parameters in the nonlinear regime. 
{This gives the physical condition for the persistence of circular 
core region in the driven system.} The characteristic core size $s$ is mainly 
determined by the dominant scale introduced either by the shear 
driving field or the smallest distance between the confining boundaries. 
Thus, in the case of highly nonlinear flow in larger domain aspect-ratio 
($L_z/L_r \gg 1$), whose scale is larger than the core size $s$, a 
part of the highly sheared boundary layer detaches from the actual boundary 
and extends deeper into the interior along the collar of core vortex. 
It begins to act as a virtual core boundary 
providing a vanishing flow velocity condition similar to the 
actual external boundary. As a consequence, the flow structure turns into a system of 
self-similar multiple co-rotating vortices having a uniform core region 
and surrounded by the corresponding highly sheared layers filled with 
small-scale weak vortices as shown in Fig.~\ref{ar-2}$(c)$. 
%
Further, we observe a steady increase in the size and strength of small-scale 
vortices with decreasing $\mu$ as shown in Fig.~\ref{ar-2}$(d)$, until the flow 
gets stabilized into more stable identical structures with its own scale ratio 
of near unity ($\approx 1:1$), which used to be $1: (\sqrt{2}-1)$ in the case 
of domain aspect-ratio of unity (AR=1)~\cite{PhysRevE.95.033204}. 

The structural changes in the flow fields can also be visualized 
from the variation in cross-section profile of the streamline potential $r\psi$ and 
the dust flow velocity profiles $(u_r,u_z)$ through the center 
of the primary vortex $(r_0,z_0)$ as shown in 
Fig.~\ref{ar-2}(e), (f) and (g) respectively. 
Here, all $r\psi$, $u_x$, and $u_y$ strengthen with the decreasing 
diffusive coefficient $\mu$. The radial shift of the center of vortex in the 
new flow structure can be observed from the extremum of the radial $r\psi$ profile 
in Fig.~\ref{ar-2}(e). The structural bifurcation 
that develops the extended virtual boundary around the core 
vortex can be demonstrated from the variation of $u_{z}(r,z_0)$ in 
Fig.~\ref{ar-2}(f). Here, $u_{z}$ at $r=r_1(\approx0)$ shows a maximum 
up to the critical value $\mu^{*}$ then begins to drop below its value 
in the interior (around the core vortex) despite that the driver strength 
always maximum at $r=r_1$. Furthermore, the radial component of the dust flow 
velocity $u_{r}(z,r_0)$ shows periodic variation that vanishes at $z\sim 0$ such 
that the line $z\sim 0$ begins to behave like the virtual boundary 
as shown in Fig.~\ref{ar-2}(g). Both the $u_x$ and $u_y$ tend to a 
saturated stable state having a uniform single shear scale between 
two equidistant opposite peaks indicating the emergence of the high-speed 
collar layer or virtual boundary (separatrix) which separate the core region 
from the highly shear boundary regions.
 
\subsection{Steady dust flow structure in the confined domain of 
aspect-ratio $L_z: L_r =3$ \label{aspect_ratio_3}}
\begin{figure}
\centering
  \includegraphics[width=0.5\textwidth]{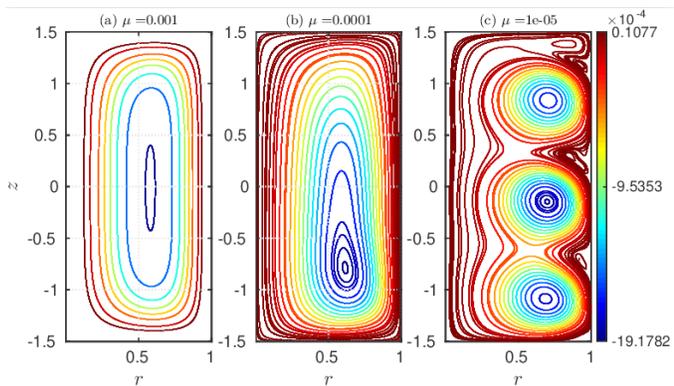}
 \caption{\small Streamlines for the steady bounded dust flow in 
 the $r$-$z$ plane of $AR = 3$ for varying 
 $(a)~\mu = 10^{-3} U_{0}L_{r}$, 
 $(b)~\mu = 10^{-4} U_{0}L_{r}$, 
 $(c)~\mu = 10^{-5} U_{0}L_{r}$ respectively 
 having fixed other system parameters 
$ \xi = 10^{-4} U_{0}/L_{r}$, $\nu =  10^{-3} U_{0}/L_{r}$.
 \label{ar-3}
}
\end{figure}

\begin{figure}
 \centering
      \includegraphics[width=0.5\textwidth]{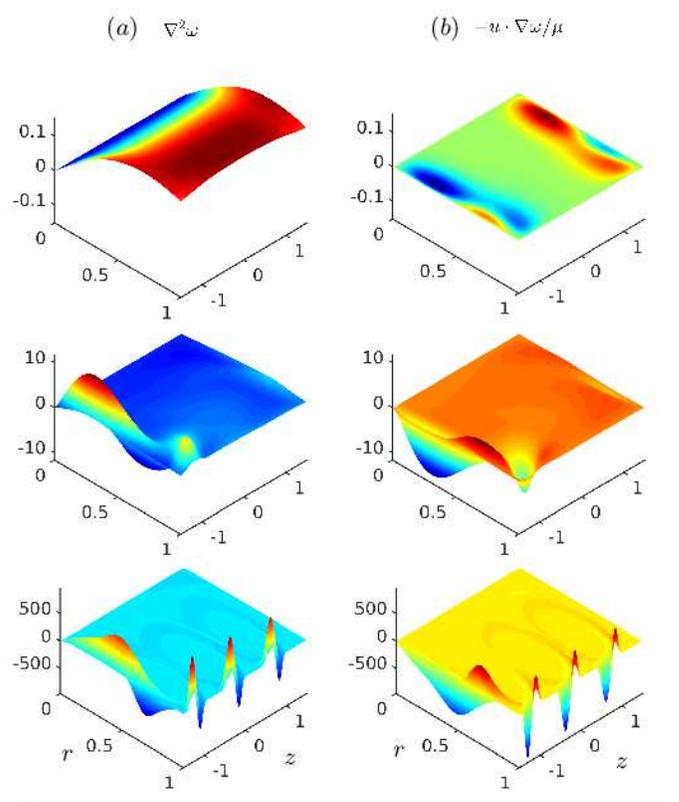}\\
\caption{\small Relative strengths of (a) the diffusive
transport and (b) the convective transport plotted from top to 
bottom for the values of dust viscosity $\mu =1\times 10^{-3} U_{0}L_{r}$, 
 $1\times 10^{-4} U_{0}L_{r}$, and $1\times 10^{-5} U_{0}L_{r}$ respectively 
 having fixed other system parameters $AR = 3$, 
 $ \xi = 10^{-4} U_{0}/L_{r}$, and 
 $\nu =  10^{-3} U_{0}/L_{r}$.
 \label{ar-3x}
}
\end{figure}

The analysis for the flow structural changes is extended further to a 
confinement domain of aspect-ratio $ L_z /L_r =3 $. The variation in the 
circulating flow structure is demonstrated with the same external 
driver $v_z$, the same system parameters $\xi$, $\nu$, and varying only the 
kinematic viscosity $\mu$ such that the corresponding effective 
Reynolds number varies from linear to highly nonlinear 
regime (Re $\simeq 0.1$ to $100$). A series of 
streamlines pattern in the axisymmetric $r$-$z$ toroidal plane 
are shown in Fig.~\ref{ar-3}. 
%
Here, as discussed in the above section~\ref{aspect_ratio_2}, the flow 
structure in the linear regime $\mu = 10^{-3} U_{0}L_{r}$ (or Re $\le 10$) 
is again symmetric and elongated circulations that closely resembles the 
geometry of the bounded domain as shown in Fig.~\ref{ar-3}$(a)$. When 
the varying parameter decreases to $\mu\approx 10^{-4}U_{0}L_{r}$ 
(or Re $\ge 20$), the vortex flow converges to a new asymmetric structure 
as shown in Fig.~\ref{ar-3}$(b)$. And further 
decrease in the parameter to $\mu~\approx 10^{-5} U_{0}L_{r}$ 
produces the qualitative topological change through the nonlinear 
structural bifurcation about a critical kinematic 
viscosity $\mu*$~\cite{PhysRevE.95.033204}. And as a consequence, 
the circulating flow pattern makes transition to a new 
self-organized steady flow structure having a sequence of three self-similar 
co-rotating circular vortices, each having a nearly uniform vorticity core 
region with a characteristic size $s$ relevant to the smaller scale of the domain 
$L_r$ ($\ll L_z$), and bounded by the shear layers 
filled with weak vortices as shown in Fig.~\ref{ar-3}$(c)$. 
In addition to this, Fig.~\ref{ar-3x} displays the corresponding changes in 
diffusive transport process $\mu\nabla^2 {\bf \omega}$ ($1^{st}$ column) relative to the 
convective process of vorticity $(\bf{u} \cdot \nabla) \bf{\omega}$ ($2^{nd}$ column) with 
decreasing the kinematic viscosity $\mu$. Diffusive transport dominant in linear regime turns 
into a balance of both convective and diffusive flow in the nonlinear regime, while the 
changes in non-conservative source and sink of vorticity are negligible. 
{It demonstrates that the changes in the steady dust fluid vortex 
structure are mainly contributed by the interplay between inertial and diffusive 
transport process} from its sources to the homogeneous 
sink present in the form of stationary background fluid.

Further, one more remarkable observation in the present analysis is that 
the bounded dust cloud retains much more momentum than any other cases 
of aspect-ratio $L_z/ L_r \ll 3$ for the same conditions of other 
dynamic factors. Therefore, the threshold parameter for the structural 
bifurcation in the present case of $L_z/L_r =3$ is found to be 
$\mu^* \approx 8\times10^{-5}U_{0}L_{r}$, whereas 
$\mu^* \approx 1\times10^{-5}U_{0}L_{r}$ for $L_z/ L_r =2$ and 
($\mu^* \approx 5\times10^{-6}U_{0}L_{r}$ in case of $L_z/L_r =0.5$). 
Moreover, the formation of a sequence of stable identical vortices are observed 
with higher $\mu^*$ in the case of larger aspect-ratios $L_{z}/L_{r}\gg 3$. 
And for the dimensions where $L_{z}/L_{r}\gg 1$ but not an integer 
multiple of $L_{r}$ say 1.5, 2.5 etc., a partially developed elongated and 
weak vortices are visible in addition to the integer 
numbers of fully grown vortices, which closely resemble the small 
vortices seen in the upper corner of Fig.~\ref{ar-2}(d) and Fig.~\ref{ar-3}(c).
{This analysis further demonstrates the condition for the existence of 
self-similar multiple co-rotating vortices in a driven-dissipative 
bounded flow system, of which the characteristic size, numbers, and strength of the 
multiple vortices depend on the aspect-ratio of the bounded flow domain.}    
In the above whole analysis, the observed scale of dust 
velocity $u\approx 10^{-3} U_{0}$ agree with the experimental observations 
of $u\sim 0.1$ - $6$ cm/sec in high Reynolds regime while shear ions are 
streaming with $c_s\sim 27$ cm/sec in various typical dust 
vortex flow experiments in both laboratory as well 
as microgravity space station~\cite{doi:10.1063/1.4929916,doi:10.1063/1.5019364,1367-2630-5-1-366}. 

\subsection{Relevant steady state co-rotating vortices in laboratory 
experiments\label{experimental_obsrv}}
Now, we discuss the correspondence between our result on multiple
co-rotating vortices and the similar observation in various recent
experiments where the physics of the transition to an apparently higher
velocity shear configuration remained not well understood or
unsupported by an adequate physics model. Here we particularly focus
on three recent examples, namely the observations reported by 
Kil-Byoung Chai {\it et~al.}~(2016)~\cite{doi:10.1063/1.4941973}, 
M. Choudhary {\it et~al.}~(2018)~\cite{doi:10.1063/1.4977454,
doi:10.1063/1.5019364}, and in one of the cases in 
biological complex fluids namely, 
Nivedita {\it et~al.}~(2017)~\cite{Nivedita}. 
All these experiments recovered the above transition in 
some form by variation of the system parameter such as 
$\xi$, $\mu$, $\nu$, and boundary
conditions in our model. 

In the experiments by 
Kil-Byoung Chai {\it et~al.}~\cite{doi:10.1063/1.4941973}, vortex motion of 
a bounded dust cloud in an rf-discharge plasma is analyzed in 
presence of various non-conservative 
ion-drag forces such as ($\nabla u_{i} \times \nabla n_{i} $) and 
($\nabla T_{i} \times \nabla n_{i} $) exerted on the dust cloud. 
As a combined effect of all sources, a system of axisymmetric toroidal 
multiple vortices including two adjacent co-rotating and counter-rotating 
vortex flows are recovered for a typical plasma density and temperature 
profile. It also observed that the multiple vortices appear only when the 
plasma density (i.e. ion density) exceeds through a critical 
value ${n_i}^*$. Furthermore, the observed secondary vortices are less 
distinct or weaker in strength than the primary vortex. 
%
From the perspective of our model, the changes in ion density $n_i$ can 
be interpreted as variation in the $\xi$ or the drive strength $U_0$, whereas 
the variation in vortex size and strength can also be understood as 
an effect of changing the aspect-ratio of 
the bounded domain as discussed in above section~\ref{aspect_ratio_3}.   

We now discuss another example of dusty plasma experiments
by M. Choudhary {\it et al.}
~\cite{doi:10.1063/1.4977454,doi:10.1063/1.5019364}
who observed steady multiple co-rotating 
vortices in an extended dust cloud confined in an inhomogeneous diffused 
rf-plasma. They first observed variation in the number of co-rotating 
vortices at the different cross-section of the extended dust cloud. 
Then they also observed the structural transition from a 
single to a state of multiple co-rotating dust vortices at a particular 
cross-section by changing the applied discharge power or voltage through 
a threshold value. In the experiment, the dissipative instability due to the 
non-zero $\nabla Q_d\times{\bf E}$ in addition to the ion drag 
forces arising from the inhomogeneous background 
plasma~\cite{doi:10.1063/1.5020416,PhysRevE.95.033204} are 
reported as the main source of vorticity. 
Here $Q_d$ is the charge on the dust particle, and ${\bf E}$ is the effective 
electric field along the ions flow direction.
Further, using the Vaulina {\it et~al.} model~\cite{1367-2630-5-1-382}, 
they interpreted the vortex characteristic size in 
terms of the diffusive scale, and its ratio to the overall dust 
cloud dimension gives the number of the sustainable vortices in the driven system.  

The explanation on the 
co-rotating character of the vortex, however, remains an open question, 
since the purely diffusive transport mechanism prescribes multiplicity
of vortex only with the spatially non-monotonic driver
~\cite{doi:10.1063/1.4929916,doi:10.1063/1.4887003,PhysRevE.91.063110})
and thus excludes the possibility of formation
of any adjacent co-rotating vortices in such system. 
In this relation, the 
explanation provided by the Choudhary {\it et~al.} although referred to 
author's work~\cite{doi:10.1063/1.4887003}, dealing with the linear 
diffusive regime but completely ignores the nonlinear convective solutions
presented in the subsequent publication~\cite{PhysRevE.95.033204}, duly relating 
them to the recovery of co-rotating vortices in flow domain of AR=1. 
Note that it is the nonlinear regime that is exactly relevant to their 
observations of adjacent self-similar co-rotating dust
vortices which are not explained by only the diffusive
transport, i.e., without including the convective nonlinear effects.
Again, from the perspective of our model, the changes in the applied 
discharge power can be modeled by varying $\xi$ or the 
drive strength $U_0$, whereas the variations in vortex size and number 
can also be understood as an effect of the aspect-ratio of the bounded 
domain as discussed in the above sections~\ref{aspect_ratio_2} 
and~\ref{aspect_ratio_3}.

For the sake of completion, and more importantly to highlight the
strong relationship between macroscopic observations in dusty plasmas
to the microscopic dynamics in complex biological fluids, 
we now discuss the observation 
by Nivedita {\it et~al.}~\cite{Nivedita} who reported the 
similar kind of structural changes and transition of the single vortex 
into multiple co-rotating vortices in various spiral 
rectangular microchannels of different aspect-ratio 
for cell sorting and micromixing devices. 
They provided systematic experimental as well as 
numerical insight into the phenomena of the structural transition by 
the variation of Dean number through a threshold value $D^*$. 
They have demonstrated the critical phenomena in term of an instability 
called Dean instability arising from the imbalance of flow pressure and 
velocity gradient near the boundary. 
In comparison to our model, pressure corresponds 
incompressible dust flow velocity ${\bf u}$, the changes in Dean number $D$ 
directly correlate to the variation of $\mu$ for the bounded flow domain 
having a fixed aspect-ratio (AR), and the Dean instability is another 
feature of the nonlinear structural bifurcation through the threshold 
parameter. Moreover, they have demonstrated the variation of the threshold 
value $D^*$ for varying the aspect-ratio, similar to the recovery of 
various critical $\mu^*$ in different aspect-ratios in  
above analysis of section~\ref{aspect_ratio_2} and section~\ref{aspect_ratio_3}.
This work actually highlights the fact that dusty plasma can be one of the 
easily realizable prototypes for the study of various driven-dissipative 
complex flow systems that can self-stabilize by making a critical 
transition to a self-similar state.
\section{Summary and Conclusions}
\label{conclusion}
We have extended and systematically analyzed the nonlinear solutions of 
the 2D hydrodynamic model for the dynamics of volumetrically driven bounded 
dust cloud in an unbounded streaming plasma~\cite{PhysRevE.95.033204} 
for different aspect-ratio (AR) of bounded domain and a wide range of the 
kinematic viscosity ($\mu$). This analysis has demonstrated the interplay between 
convective and diffusive transport processes in determining the steady vortex 
structure of the bounded dust cloud for the wide range of Reynolds numbers 
from linear to a highly nonlinear regime. 
This analysis also gives the insight that the effective 
Reynolds number of the bounded flow system with fixed source and sink of 
momentum can be controlled through 
the aspect-ratio of the flow domain in addition to the fluid's diffusive 
coefficient $\mu$. 
Therefore, the vortex structural changes from the linear to 
highly nonlinear flow regimes are analyzed in a larger domain of AR=2 and AR=3. 
The series of steady flow structure in case of AR=2, has clearly demonstrated the 
nonlinear effects of structural changes from symmetric into asymmetric nature in the 
nonlinear regime that exhibits regions of localized acceleration in 
the flow field. Moreover, it reveals the physical reason 
for the flow trajectory turns into more circular and the nonlinear 
structural bifurcation takes place about a threshold parameter $\mu^*$. Thus the 
new vortex structure in highly nonlinear regime is characterized by the critical 
transition into new self-similar multiple co-rotating vortices having  
circular core region of single characteristics size $s$ on 
which the viscous dissipations is negligible (uniform vorticity) and 
hence it would persist for a long range of system parameters in the nonlinear 
regime. It has further revealed that the size $s$ is mainly determined 
by the dominant scale introduced either by the shear driving 
field or the smallest distance between the confining boundary. Thus, for the 
cases of ($AR\gg 1$), the multiple co-rotating vortices get stabilized with 
a scale ratio of near unity ($\approx 1:1$), which were scale ratio 
of $1: (\sqrt{2}-1)$ in case of aspect-ratio of unity (AR=1)~\cite{PhysRevE.95.033204}.
Further, the series of the steady flow structure in case of AR=3 displays clearly 
how the structure changes with varying the flow domain's aspect-ratio and 
kinematic viscosity. 
Thus, the driven flow system allows to retains more momentum either 
through an increase in $AR$ or decreasing $\mu$. As a consequence, the threshold 
parameter $\mu^*$ for the structural bifurcation decrease in the 
bounded domain of larger $AR$. Additionally, this work also demonstrated the impact 
of the system parameters in determining the characteristic size, shape, strength, 
and numbers of the self-similar co-rotating vortices in many of relevant 
driven-dissipative flow systems. 

The whole analysis of unusual structural dependence on the dynamical parameters 
including the AR of the flow domain is relevant to many of recent dusty plasma 
experiments as well as biological complex fluid flows~\cite{doi:10.1063/1.4941973,
Nivedita,doi:10.1063/1.5019364}. This signifies the fact that the proposed structural 
bifurcation through the threshold parameter is another aspect of the Dean instability 
or dissipative instability reported in the real experimental systems. 
This nonlinear threshold phenomenon may associate with shear instabilities which 
can be investigated further by extending the same analysis to real-time dependence 
in well-defined parameter regimes. 

\section{Acknowledgments}
Author L. Modhuchandra acknowledge Prof. Abhijit Sen and late Prof. P. K. Kaw, 
Institute for Plasma Research, India, for the invaluable supports, encouragements, 
scientific suggestions, and criticisms. The research was supported by the State 
Administration of Foreign Experts Affairs - Foreign Talented Youth Introduction 
Plan Grant No. WQ2017ZGKX065, the National Magnetic Confinement Fusion Program of 
China under Grant Nos. 2014GB124002 and 2015GB101004. 
Author P. Zhu also acknowledges the support from U.S. DOE 
grant Nos. DE-FG02-86ER53218 and DE-FC02-08ER54975.
The work used the resources of Supercomputing Center of 
University of Science and Technology of China. 
%
\end{document}